# Analytical study of spherical cloak/anti-cloak interactions

Giuseppe Castaldi [a], Ilaria Gallina [a], Vincenzo Galdi [a,*], Andrea Alù [b], and Nader Engheta [c]

[a] Waves Group, Department of Engineering, University of Sannio, Corso Garibaldi 107, I-82100, Benevento, Italy

[b] Department of Electrical and Computer Engineering, The University of Texas at Austin, 1 University Station, C0803, Austin, TX 78712, USA

[c] Department of Electrical and Systems Engineering, University of Pennsylvania, 200 South 33rd Street, Room 215, Moore Building, Philadelphia, PA 19104, USA

**ABSTRACT** - The intriguing concept of "anti-cloaking" has been recently introduced within the framework of transformation optics (TO), first as a "countermeasure" to invisibility-cloaking (i.e., to restore the scattering response of a cloaked target), and more recently in connection with "sensor invisibility" (i.e., to strongly reduce the scattering response while maintaining the field-sensing capabilities). In this paper, we extend our previous studies, which were limited to a two-dimensional *cylindrical* scenario, to the three-dimensional *spherical* case. More specifically, via a generalized (coordinate-mapped) Mie-series approach, we derive a general analytical full-wave solution pertaining to plane-wave-excited configurations featuring a spherical object surrounded by a TO-based invisibility cloak coupled via a vacuum layer to an anti-cloak, and explore the various interactions of interest. With a number of selected examples, we illustrate the cloaking and field-restoring capabilities of various configurations, highlighting similarities and differences with respect to the cylindrical case, with special emphasis on sensor-cloaking scenarios and ideas for approximate implementations that require the use of *double-positive* media only.

[*] Corresponding author. Tel: +39 0824-305809, Fax: +39 0824-305840

E-mail addresses: castaldi@unisannio.it (G. Castaldi), ilaria.gallina@unisannio.it (I. Gallina), vgaldi@unisannio.it (V. Galdi), alu@mail.utexas.edu (A. Alù), engheta@ee.upenn.edu (N. Engheta).

# 1. Introduction

During the past few years, there has been an increasingly renewed interest in the development of strategies for achieving *invisibility* ("cloaking") of objects to waves of various nature (electromagnetic, acoustic, biharmonic, etc.). Besides the fascinating mathematical and physical aspects of the phenomenological underpinnings, such interest is motivated by the potentially disruptive technological implications and applications to a vast variety of fields, ranging from camouflaging to non-invasive medical diagnostics.

In this framework, a broad distinction may be made between *active* and *passive* strategies. The first class of strategies is essentially based on the use of active *sources* for creating *destructive interference* in the region to be cloaked and reconstructing the signal elsewhere (see, e .g., [1-3]), in a way that somehow resembles standard noise-cancellation techniques [4]. Conversely, the second class of strategies is based on the design of suitable *material* structures that, when placed around or nearby the object, may drastically suppress its near/far-field visibility (scattering). With specific reference to the electromagnetic (EM) case, prominent approaches belonging to this class include the scattering cancellation [5-7], anomalous localized resonances [8], transformation optics (TO) [9-15], inverse design optical elements [16], transmission lines [17,18], and mantle cloak [19]. The reader is referred to [20,21] for recent comparative studies.

In what follows, we will be mainly concerned with the TO-based approach, in which a suitable curved coordinate transformation (from a fictitious auxiliary space to the actual physical space) is designed so as to create a "hole," i.e., a region *inaccessible* to EM waves wherein an object can be concealed. Thanks to the co-variance properties of Maxwell's equations, the *energy re-routing* effects of such coordinate mapping can be equivalently obtained in a *flat*, Cartesian space filled by a suitable (inhomogeneous, anisotropic) "transformation medium," whose constitutive properties can be systematically derived from the Jacobian matrix of the transformation (see [9-15] for more details).

While the above TO-based cloaking concept can be intuitively understood in terms of the coordinate-mapping-induced *ray bending* properties of the transformation medium, its validity is by no means restricted to the asymptotic (high-frequency) ray-optical regime. In fact, it was shown in a number of analytical full-wave studies on canonical (cylindrical, spherical) geometries [22-25] that, in the *ideal* limit (implying a lossless, non-dispersive, extreme-parameter transformation medium) a TO-based cloak generally provides *complete isolation* (i.e., no power exchange) between the inside and outside regions, at *any* frequency. Nevertheless, for the two-dimensional (2-D) cylindrical scenario, it was shown by Chen *et al.* [26] that, if the cloaked region contains a *complementary-designed* transformation medium, the cloaking effect can, in principle, be totally or partially *compensated*. Such effect was accordingly referred to as "anti-cloaking," and it was shown that a *double-negative* (DNG) transformation medium (i.e., with both permittivity and permeability parameters negative [27]) was generally required for its implementation. Expanding upon this concept, we explored in [28] a more general scenario, with cloak and anti-cloak not necessarily contiguous, and the anti-cloak not necessarily DNG. From this study, it emerged that, besides the originally proposed cloaking "countermeasure" [26], the anti-cloak concept could also find interesting applications to creating a *multiply-connected* cloaked region while still maintaining the possibility of sensing the outside field. In a more recent paper [29], we revisited this concept, and contextualized it within the emerging framework of "sensor invisibility" that had been in parallel introduced [30] and explored in various configurations [31,32].

In this paper, we extend the above studies to the 3-D *spherical* scenario, of interest for more realistic applications. As for the cylindrical case, we use an analytical full-wave approach, based on the generalized Mie theory developed in [24,33]. Besides the inherent formal complications, our study reveals some interesting aspects and effects not emerged in our previous cylindrical studies, and also focuses on ideas for the implementation of approximate anti-cloaking effects which involve only *double-positive* (DPS) media (i.e., with both permittivity and permeability parameters positive). Such ideas for implementation, already successfully explored in 2-D square configurations [34], would remove the most significant fabrication challenges.

Accordingly, the rest of the paper is laid out as follows. In Section 2, we introduce the problem geometry and outline its formulation. In Section 3, with reference to the time-harmonic plane-wave scattering, we illustrate the general analytical solution, and subsequently particularize it to the various scenarios of interest. In Section 4, we present and discuss some representative numerical results. Finally, in Section 5, we provide some brief concluding remarks and hints for future research.

## 2. Problem geometry and formulation

In parallel with [28,29], we start from a fictitious auxiliary space $(x',y',z')$ (illustrated in Fig. 1(a)), containing a two-layer piecewise-homogeneous, isotropic spherical configuration, immersed in vacuum, which, in the associated spherical $(r',\theta',\phi')$ reference system, may be parameterized by the permittivity and permeability distributions

$$\varepsilon'(r') = \begin{cases} \varepsilon_1, & 0 < r' < R_1 \\ \varepsilon_2, & R_1 < r' < R_2 \\ \varepsilon_0, & r' > R_2 \end{cases}, \quad \mu'(r') = \begin{cases} \mu_1, & 0 < r' < R_1 \\ \mu_2, & R_1 < r' < R_2 \\ \mu_0, & r' > R_2 \end{cases}, \tag{1}$$

where $\varepsilon_0$ and $\mu_0$ denote the vacuum dielectric permittivity and magnetic permeability, respectively. By applying to this configuration the piecewise linear radial (in the associated $(r,\theta,\phi)$ spherical reference system) coordinate transformation (see Fig. 1(b))

$$r' = f(r) = \begin{cases} r, & r < R_1, \; r > R_4, \\ R_1'\left(\dfrac{R_2 + \Delta_2 - r}{R_2 + \Delta_2 - R_1}\right), & R_1 < r < R_2, \\ R_4\left(\dfrac{r - R_3 + \Delta_3}{R_4 - R_3 + \Delta_3}\right), & R_3 < r < R_4, \end{cases} \qquad (2)$$

we finally obtain a four-layer spherical configuration in the actual physical space $(x, y, z)$ (see Fig. 1(c)). Such configuration constitutes the 3-D spherical generalization of the cylindrical scenario in [28,29], and comprises a homogeneous, isotropic sphere (of radius $R_1$ and constitutive parameters $\varepsilon_1, \mu_1$) surrounded by an anti-cloak shell $(R_1 < r < R_2)$ with constitutive parameters *opposite in sign* to $\varepsilon_2, \mu_2$, a vacuum gap $(R_2 < r < R_3)$, and an outermost cloak shell $(R_3 < r < R_4)$, all immersed in vacuum. The cloak and anti-cloak constitutive-tensor-components (in spherical coordinates) may systematically be derived from (1) and (2) as (see [11,24,33] for details)

$$\varepsilon_r(r) = \frac{\varepsilon'(r')}{\dot{f}(r)}\left(\frac{r'}{r}\right)^2, \quad \varepsilon_\theta(r) = \varepsilon_\phi(r) = \varepsilon'(r')\dot{f}(r),$$
$$\mu_r(r) = \frac{\mu'(r')}{\dot{f}(r)}\left(\frac{r'}{r}\right)^2, \quad \mu_\theta(r) = \mu_\phi(r) = \mu'(r')\dot{f}(r). \qquad (3)$$

Here, and henceforth, an overdot indicates differentiation with respect to the argument. Moreover, as in [28,29], the vanishingly small parameters $\Delta_2$ and $\Delta_3$ in (2) parameterize the departure from the *ideal* (i.e., no parameter truncation) configurations.

Note that our configuration here differs from that in [26] for the possible presence of the vacuum gap separating the cloak and anti-cloak, and also for the constitutive properties of the fictitious auxiliary space which, in our case, yield an *intrinsically matched* anti-cloak, not aimed at restoring the scattering response of the inner sphere.

As for the cylindrical case in [28,29], we expect to be able to suitably tailor (via the small parameters $\Delta_2$ and $\Delta_3$) the competing cloak/anti-cloak effects so as to create an *effectively cloaked* region in the vacuum gap $R_2 < r < R_3$, while still being able to restore a non-negligible field in the inner spherical region $r < R_1$. However, by comparison with the cylindrical scenario in [28,29], some differences appear in the presence of the additional parameters $\varepsilon_2, \mu_2, R_1'$. In particular, it can be observed from (2) and Fig. 1(b) that a choice $R'_1 \neq R_1$ renders the coordinate transformation in (2) *inherently discontinuous*. With the exception of only one case (see Section 3.3 below) where such parameters will be exploited as extra degrees of freedom, we will be assuming $\varepsilon_2 = \varepsilon_1, \mu_2 = \mu_1, R'_1 = R_1$, thereby recovering the formal structure in [28,29]. Another important difference is in the variation ranges of the constitutive parameters. For the cloak shell (always DPS), it can readily be shown that

$$\frac{R_4 \Delta_3^2 \varepsilon_0}{R_3^2 (R_4 - R_3)} < \varepsilon_r < \frac{(R_4 - R_3)\varepsilon_0}{R_4}, \quad \varepsilon_{\theta,\phi} = \frac{R_4 \varepsilon_0}{R_4 - R_3}, \quad R_3 < r < R_4, \tag{4}$$

and therefore only the radial component is inhomogeneous, ranging from *near-zero* (as $\Delta_3 \to 0$) to *finite* values. Formally analogous results (by replacing $\varepsilon$ with $\mu$) are obtained for the permeability components. Similar considerations hold in absolute-value for the anti-cloak shell too, viz.,

$$\frac{R'_1 \Delta_2^2 |\varepsilon_2|}{R_2^2 (R_2 - R_1)} < |\varepsilon_r| < \frac{(R_2 - R_1)|\varepsilon_2|}{R'_1}, \quad |\varepsilon_{\theta,\phi}| = \frac{R'_1 |\varepsilon_2|}{(R_2 - R_1)}, \quad R_1 < r < R_2, \tag{5}$$

but the sign of the constitutive parameters depend on those of $\varepsilon_2, \mu_2$.

In what follows, we study the EM response of the four-layer spherical cloak/anti-cloak configuration in Fig. 1(c) to a time-harmonic $[\exp(-i\omega t)]$ plane wave with unit-amplitude $x$-directed electric field, impinging from the positive $z$-direction,

$$\boldsymbol{E}^i(z) = \exp(ik_0 z)\hat{\boldsymbol{u}}_x, \tag{6}$$

where $k_0 = \omega\sqrt{\varepsilon_0 \mu_0} = 2\pi/\lambda_0$ denotes the vacuum wavenumber (with $\lambda_0$ being the corresponding wavelength). In (6), and henceforth, boldface symbols identify vector quantities, and $\hat{\boldsymbol{u}}_\alpha$ denotes an $\alpha$-directed unit vector.

## 3. Analytical derivations

*3.1 General solution*

Our general analytical full-wave solution is based on the generalized (coordinate-mapped) Mie-series approach proposed in [24,33]. First, it is expedient to split the impinging plane wave in (6) as the sum of transverse electric (TE) and magnetic (TM) components,

$$\boldsymbol{E}^i(z) = \boldsymbol{E}^i_{TE}(r,\theta,\phi) + \boldsymbol{E}^i_{TM}(r,\theta,\phi), \tag{7}$$

which can be in turn represented in terms of scalar Debye potentials:

$$\begin{aligned}\boldsymbol{E}^i_{TE} &= -\varepsilon_0^{-1}\nabla\times(\hat{\boldsymbol{u}}_r \Phi^i_{TE}), & \boldsymbol{H}^i_{TE} &= -i\omega^{-1}\mu_0^{-1}\nabla\times\boldsymbol{E}^i_{TE}, \\ \boldsymbol{H}^i_{TM} &= \mu_0^{-1}\nabla\times(\hat{\boldsymbol{u}}_r \Phi^i_{TM}), & \boldsymbol{E}^i_{TM} &= i\omega^{-1}\varepsilon_0^{-1}\nabla\times\boldsymbol{H}^i_{TM},\end{aligned} \tag{8}$$

admitting the following spherical harmonic expansion [35, Chap. 6]:

$$\Phi_{TE}^i(r,\theta,\phi) = \sum_{n=1}^{\infty} i^n \frac{2n+1}{n(n+1)} \sin\phi P_n^1(\cos\theta) S_n(k_0 r),$$

$$\Phi_{TM}^i(r,\theta,\phi) = \sum_{n=1}^{\infty} i^n \frac{2n+1}{n(n+1)} \cos\phi P_n^1(\cos\theta) S_n(k_0 r).$$

(9)

In (9), $P_n^m$ denote the associated Legendre functions of the first kind [36, Section 8], and $S_n(x) = x j_n(x)$ denote the $n$th-order Riccati-Bessel functions of the first kind [36, Section 10.3], with $j_n$ denoting the $n$th-order spherical Bessel function of the first kind [36, Section 10.1].

The *total* fields in the various regions of Fig. 1(c) may be likewise represented in terms of scalar Debye potentials in the following compact form:

$$\boldsymbol{E}_{TE}^{(\nu)} = -\underline{\underline{\varepsilon}}^{-1} \cdot \nabla \times \left( \dot{f} \hat{\boldsymbol{u}}_r \Phi_{TE}^{(\nu)} \right), \quad \boldsymbol{H}_{TE}^{(\nu)} = -i\omega^{-1} \underline{\underline{\mu}}^{-1} \cdot \nabla \times \boldsymbol{E}_{TE}^{(\nu)},$$

$$\boldsymbol{H}_{TM}^{(\nu)} = \underline{\underline{\mu}}^{-1} \cdot \nabla \times \left( \dot{f} \hat{\boldsymbol{u}}_r \Phi_{TM}^{(\nu)} \right), \quad \boldsymbol{E}_{TM}^{(\nu)} = i\omega^{-1} \underline{\underline{\varepsilon}}^{-1} \cdot \nabla \times \boldsymbol{H}_{TM}^{(\nu)},$$

$$R_{\nu-1} < r < R_\nu, \quad \nu = 1,..,5,$$

(10)

where the "dummy" parameters $R_0 = 0$ and $R_5 = \infty$ have been introduced for notational convenience. The constitutive tensors $\underline{\underline{\varepsilon}}$ and $\underline{\underline{\mu}}$ are given in (3) for the cloak and anti-cloak layers, and reduce to the appropriate scalar quantities in the vacuum ($R_2 < r < R_3, r > R_4$) and dielectric target ($r < R_1$) regions. Following [24,33], the scalar Debye potentials in (10) can be expanded by mapping (via (2)) the straightforward Mie-series solution in the fictitious auxiliary space (plane-wave scattering by a two-layer piecewise-homogeneous, isotropic spherical configuration, cf. Fig. 1(a)), viz.,

$$\Phi_{TE}^{(\nu)}(r,\theta,\phi) = \sum_{n=1}^{\infty} i^n \frac{2n+1}{n(n+1)} \left\{ \left( a_{n,TE}^{(\nu)} + \delta_{\nu 5} \right) S_n[g(r)] + b_{n,TE}^{(\nu)} C_n[g(r)] \right\} \sin\phi P_n^1(\cos\theta),$$

$$\Phi_{TM}^{(\nu)}(r,\theta,\phi) = \sum_{n=1}^{\infty} i^n \frac{2n+1}{n(n+1)} \left\{ \left( a_{n,TM}^{(\nu)} + \delta_{\nu 5} \right) S_n[g(r)] + b_{n,TM}^{(\nu)} C_n[g(r)] \right\} \cos\phi P_n^1(\cos\theta),$$

(11)

$$R_{\nu-1} < r < R_\nu, \quad \nu = 1,..,5,$$

where $C_n(x) = -x y_n(x)$ denote the $n$th-order Riccati-Bessel functions of the second kind [36, Section 10.3], with $y_n$ denoting the $n$th-order spherical Bessel functions of the second kind [36, Section 10.1], and

$$g(r) = \begin{cases} k_0 r, & R_2 < r < R_3, \\ \omega \sqrt{\varepsilon'[f(r)] \mu'[f(r)]} f(r), & r < R_2, r > R_3. \end{cases}$$

(12)

In (11), the Kronecker delta $\delta_{\nu 5}$ accounts for the presence of the incident field (cf. (9)) in the exterior vacuum region $r > R_4$, while the unknown expansion coefficients $a_{n,TE}^{(\nu)}, b_{n,TE}^{(\nu)}, a_{n,TM}^{(\nu)}, b_{n,TM}^{(\nu)}$ need to be computed by enforcing the continuity/boundary conditions. In particular, one finds $b_n^{(1)} = 0$ (field-finiteness condition at $r = 0$) and $b_n^{(5)} = -i a_n^{(5)}$ (radiation condition), irrespective of the TE/TM polarization. The tangential-field-continuity conditions at the four interfaces yield the remaining 16 (numerable infinities of) unknown coefficients in terms of two *decoupled* (TE/TM) 8×8 linear systems of equations, whose straightforward but cumbersome solutions are not reported here for brevity. Instead, as in [29], we generally focus on their *scaling laws* (for $\Delta_2, \Delta_3 \to 0$), via the use of the asymptotic Landau notation $O(\cdot)$, neglecting irrelevant constants and higher-order terms. In what follows, for notational convenience, we also omit the TE/TM subfixes in those results that do not depend on the polarization state.

*3.2 Standard configuration*

3.2.1   Vacuum gap: Cloaked region and imperfect field recovery

We start considering the case
$$\varepsilon_2 = \varepsilon_1, \mu_2 = \mu_1, R_1' = R_1, \tag{13}$$
which represents the *direct* generalization of the cylindrical scenario in [28]. In this case, the tangential-field continuity conditions at the interfaces $r = R_1$ and $r = R_4$ straightforwardly yield
$$a_n^{(1)} = a_n^{(2)}, \ b_n^{(2)} = 0, \tag{14}$$
$$a_n^{(5)} = i b_n^{(4)}, \ a_n^{(4)} = 1 + i b_n^{(4)}, \tag{15}$$
whereas those at the remaining interfaces $r = R_2$ and $r = R_3$ are trickier, in view of the *singular* behavior exhibited by the Riccati-Bessel functions of the second kind in the expansion (11) in the limit as $\Delta_2, \Delta_3 \to 0$ (i.e., ideal cloak and anti-cloak). Recalling the small-argument approximations of the Riccati-Bessel functions (cf. Eqs. (9.1.7) and (9.1.9) in [36]),
$$S_n(x) \sim \frac{\pi^{1/2} x^{n+1}}{2^{n+1} \Gamma\left(n + \frac{3}{2}\right)}, \quad C_n(x) \sim -\frac{\Gamma\left(n + \frac{1}{2}\right)}{\pi^{1/2} 2^n x^n}, \quad |x| \ll 1, n \geq 1, \tag{16}$$
with $\Gamma$ denoting the Gamma function [36, Section 6.1], we obtain
$$a_n^{(1,2)} \sim O\left(\frac{\Delta_3^{n+1}}{\Delta_2^n}\right), \tag{17}$$

$$a_n^{(3)} \sim b_n^{(3)} \sim O\left(\Delta_3^{n+1}\right), \tag{18}$$

$$a_n^{(5)} = ib_n^{(4,5)} \sim O\left(\Delta_3^{2n+1}\right). \tag{19}$$

Therefore, as for the cylindrical case [28,29], we find that for vanishingly small $\Delta_3$ the coefficients pertaining to the field transmitted in the vacuum gap $R_2 < r < R_3$ (cf. (18)) and to that scattered in the exterior region $r > R_4$ (cf. (19)) decay *algebraically*, and hence these fields are *strongly suppressed*. Conversely, the coefficients pertaining to the field transmitted in the inner spherical region $r < R_1$ depend on the ratio $\Delta_3/\Delta_2$, and thus they may be in principle recovered. In particular, unlike the cylindrical case [28] (for which the field recovery was achieved by letting $\Delta_2, \Delta_3 \to 0$ while keeping their ratio *finite*), here we need $\Delta_2$ to go to zero *faster* than $\Delta_3$. More specifically, assuming $\Delta_2 \sim O\left(\Delta_3^q\right)$, we obtain from (17)

$$a_n^{(1)} \sim O\left(\Delta_3^{n+1-qn}\right), \tag{20}$$

from which it is clear that for $q \leq 1$ all the coefficients are asymptotically vanishing, whereas we can recover the orders $n > 1/(q-1)$ for $1 < q < 2$, and all the orders for $q \geq 2$. In these last two cases, it should be noted that, although the coefficients grow algebraically as $n \to \infty$, the series in (11) are still summable in view of the exponentially decaying asymptotic (large-order) behavior of the Riccati-Bessel functions (cf. Eq. (9.3.1) in [36]),

$$\lim_{n \to \infty} a_n^{(1)} S_n\left[k_0 g(r)\right] = \lim_{n \to \infty} O\left[\Delta_3^{1+n(1-q)} n^{-n}\right] = 0. \tag{21}$$

To sum up, similar to the cylindrical case (though via a different scaling law of the parameters $\Delta_2$ and $\Delta_3$), it is still possible to restore in the inner spherical region $r < R_1$ a *distorted* version of the impinging field, while maintaining the vacuum gap $R_2 < r < R_3$ *effectively cloaked*, with very weak exterior scattering.

3.2.2 No vacuum gap: Perfect field restoration and sensor cloaking

It is also interesting to consider the case in the absence of the vacuum gap (i.e., $R_3 = R_2$). Particularizing the general solution in (10) and (11) to the above configuration, and proceedings as in Section 3.2.1, we obtain

$$a_n^{(1,2)} \sim \frac{O\left(\frac{\Delta_3^n}{\Delta_2^n}\right)}{1+O\left(\frac{\Delta_2}{\Delta_3}\right)}, \tag{22}$$

$$a_n^{(5)} = ib_n^{(4,5)} \sim O\left(\Delta_3^{2n+1}\right)\left[\frac{1+O\left(\frac{\Delta_3}{\Delta_2}\right)}{1+O\left(\frac{\Delta_3}{\Delta_2}\right)}\right]. \tag{23}$$

From (22), it is clear that it is now possible to recover, in principle, *all* the coefficient pertaining to the field transmitted in the inner spherical region $r < R_1$ by letting $\Delta_3, \Delta_2 \to 0$ while keeping their ratio *finite*. In particular, looking at the actual expression of the coefficients, we obtain for the TM case

$$a_{n,TM}^{(1,2)} \sim \frac{\varepsilon_1\mu_1 k_0(1+2n)(R_2-R_1)R_4}{\mu_0\varepsilon_1 n k_1(R_4-R_3)R_1\frac{\Delta_2}{\Delta_3}+k_0(R_2-R_1)R_4(1+n)\sqrt{\varepsilon_1\mu_1\varepsilon_0\mu_0}}\left[\frac{k_0R_4(R_2-R_1)\Delta_3}{k_1R_1(R_4-R_3)\Delta_2}\right]^n, \tag{24}$$

with $k_1 = \omega\sqrt{\varepsilon_1\mu_1}$ denoting the ambient wavenumber in the inner spherical region, while the dual TE case can be simply obtained by interchanging the permittivities and permeabilities. Interestingly, the particular choice

$$\frac{\Delta_3}{\Delta_2} = \frac{\sqrt{\varepsilon_1\mu_1}R_1(R_4-R_3)}{\sqrt{\varepsilon_0\mu_0}R_4(R_2-R_1)}, \quad \varepsilon_0\mu_1 = \varepsilon_1\mu_0 \tag{25}$$

yields

$$a_n^{(1,2)} \sim \sqrt{\frac{\varepsilon_1\mu_1}{\varepsilon_0\mu_0}} = \frac{\varepsilon_1}{\varepsilon_0}, \tag{26}$$

which implies that the impinging plane wave can be *perfectly* restored in the inner spherical region $r < R_1$, while maintaining a very weak exterior scattering. Moreover, exploiting (26), we can actually control the amplitude of the restored field via the ratio $\varepsilon_1/\varepsilon_0$. This last effect turns out to be peculiar of the spherical scenario, with no counterpart in the cylindrical case [28,29].

A particularly interesting application of the above configuration is in the framework of "sensor cloaking" [29-32], where one is interested in strongly suppressing the visibility (scattering) of a sensor, while maintaining its field-sensing (absorption) capabilities. Such possibility was recently introduced in [30], based on the scattering-cancellation approach, and subsequently pursued via alternative approaches based on Fano resonances [31] and TO [29,32].

In [29], in connection with the cylindrical scenario, we showed that a suitable tailoring of the competing cloak/anti-cloak effects could provide an effective TO-based route to sensor cloaking. For that scenario, we proved analytically (and verified numerically) that the best performance was indeed obtained in the absence of the vacuum gap [29]. Although the generalization of such conclusion to the spherical case of interest here cannot be taken for granted and would be rather involved to address, we will show hereafter (see Section 4 below) that reasonably good performance can be obtained by considering a configuration featuring a slightly lossy spherical particle (mimicking the sensor loading effects) surrounded by a cloak/anti-cloak configuration with no vacuum gap.

*3.3 Approximate DPS implementation*

As previously highlighted, the constitutive parameters characterizing the anti-cloak are opposite in sign to those possessed by the medium filling the auxiliary-space region to be transformed. Based on this observation, we showed in [28] that cylindrical anti-cloaking effects may also be obtained via DPS transformation media, by transferring the DNG character to the inner region. This somehow relaxes some of the practical feasibility limitations, since the inner region is homogeneous and characterized by finite-value parameters. Nevertheless, an implementation that avoided the use of *any* DNG materials would be much more convenient for practical applications. In [34], we explored, with promising outcomes, the idea for an approximate anti-cloak implementation based entirely on DPS materials, in connection with a particular class of 2-D transformations similar to those introduced in [37] to design a square-shaped invisibility cloak. In what follows, we explore such possibility in the spherical scenario of interest.

Our approach is based on the judicious exploitation of the additional parameters $\varepsilon_2, \mu_2, R'_1$ purposely introduced in the model. The basic idea is to enforce the DPS character of the anti-cloak by starting with a DNG material in the auxiliary-space region $R_1 < r' < R_2$ (Fig. 1 (a)), i.e., choosing $\varepsilon_2 < 0, \mu_2 < 0$. In view of the *decreasing* character (i.e., $\dot{f} < 0$) of the anti-cloak coordinate-transformation (see. Fig. 1(b)), it is readily understood that the corresponding transformation-medium (cf. (3)) will be DPS. Moreover, in order to maintain the DPS character of the inner spherical region $r < R_1$, we need to keep $\varepsilon_1 > 0, \mu_1 > 0$. This inherently implies a parameter *mismatch* at the interface $r = R_1$, which would deteriorate the anti-cloaking function. In order to (at least partially) *compensate* this mismatch, we purposely render the coordinate transformation *discontinuous* at that interface, by choosing $R'_1 \neq R_1$ (see Fig. 1(b)). Accordingly, indicating with $k_2 = \omega\sqrt{\varepsilon_2 \mu_2}$ the ambient wavenumber in the auxiliary-space region $R_1 < r' < R_2$, the tangential-field continuity conditions at the interface $r = R_1$ now yield, for the TM case,

$$a_{n,TM}^{(2)} = \frac{\mu_2 \left[ k_1 \varepsilon_2 \dot{S}_n(k_1 R_1) C_n(k_2 R_1') - \varepsilon_1 k_2 S_n(k_1 R_1) \dot{C}_n(k_2 R_1') \right]}{k_2 \varepsilon_1 \mu_1 \left[ \dot{S}_n(k_2 R_1') C_n(k_2 R_1') - S_n(k_2 R_1') \dot{C}_n(k_2 R_1') \right]} a_{n,TM}^{(1)}, \quad (27)$$

$$b_{n,TM}^{(2)} = \frac{\mu_2 \left[ -\varepsilon_1 k_2 S_n(k_1 R_1) \dot{S}_n(k_2 R_1') + k_1 \varepsilon_2 \dot{S}_n(k_1 R_1) S_n(k_2 R_1') \right]}{k_2 \varepsilon_1 \mu_1 \left[ \dot{S}_n(k_2 R_1') C_n(k_2 R_1') - S_n(k_2 R_1') \dot{C}_n(k_2 R_1') \right]} a_{n,TM}^{(1)}, \quad (28)$$

instead of (14), while the conditions in (15) still hold. Again, the dual conditions pertaining to the TE polarization can readily be derived from (27) and (28) by interchanging the permittivies and permeabilities. As it will be clear below, instrumental to the anti-cloaking effects is the condition $b_n^{(2)} = 0$, which can be re-enforced in (28), yielding

$$\frac{\dot{S}_n(k_2 R_1')}{S_n(k_2 R_1')} = \frac{-|\varepsilon_2|\sqrt{\varepsilon_1 \mu_1} \dot{S}_n(k_1 R_1)}{\varepsilon_1 \sqrt{\varepsilon_2 \mu_2} S_n(k_1 R_1)}. \quad (29)$$

In what follows, we assume $\varepsilon_2/\varepsilon_1 = \mu_2/\mu_1$, so that the conditions in (29) are valid for both TE and TM polarizations. While it is impossible (for $\varepsilon_2 \neq \varepsilon_1$ and $\mu_2 \neq \mu_1$) to exactly satisfy the above numerable infinity of transcendental equations, *approximate* anti-cloaking effects may still be obtained by enforcing only a limited number of them (e.g., those pertaining to the lowest $n$ orders) by exploiting the available free (though sign-constrained) parameters (e.g., $R_1' > 0, \varepsilon_1 > 0, \mu_1 > 0, \varepsilon_2 < 0, \mu_2 < 0$). In order to highlight the key role played by the conditions in (29) (i.e., $b_n^{(2)} = 0$), we consider possible imperfect solutions in the form

$$\dot{S}_n(k_2 R_1') = -S_n(k_2 R_1') \frac{|\varepsilon_2|\sqrt{\varepsilon_1 \mu_1} \dot{S}_n(k_1 R_1)}{\varepsilon_1 \sqrt{\varepsilon_2 \mu_2} S_n(k_1 R_1)} + \Delta_n, \quad (30)$$

with $\Delta_n$ parameterizing the mismatch. By proceeding as in Section 3.2.2, and expanding via a first-order Taylor expansion the dependence on $\Delta_n$, it can be shown that, in the limit as $\Delta_n \to 0$,

$$a_n^{(1,2)} \sim \frac{O(\Delta_3^{n+1} \Delta_2^{n+1})}{O(\Delta_2^{2n+1}) + O(\Delta_n)}, \quad (31)$$

$$a_n^{(3)} \sim b_n^{(3)} \sim O(\Delta_3^{n+1}) \left[ \frac{O(\Delta_2^{2n+1}) + O(\Delta_n)}{O(\Delta_2^{2n+1}) + O(\Delta_n)} \right], \quad (32)$$

$$a_n^{(5)} = i b_n^{(4,5)} \sim O(\Delta_3^{2n+1}) \left[ \frac{O(\Delta_2^{2n+1}) + O(\Delta_n)}{O(\Delta_2^{2n+1}) + O(\Delta_n)} \right], \quad (33)$$

from which the vanishingly small character of the field transmitted in the vacuum gap and the scattered field is evident. For the field transmitted in the inner spherical region, by comparing (31) with (17), we observe that, for a given order $n$, similar considerations as for the standard implementation (cf. (20) and related discussion) hold, provided that $\Delta_n \to 0$ at least as fast as $\Delta_2$. Conversely, for *finite* values of $\Delta_n$, it can be shown that, while (18) and (19) still hold, the field transmitted in the inner spherical region can no longer be recovered since the corresponding coefficients now *decay algebraically*, viz.,

$$a_n^{(1)} \sim O\left(\Delta_3^{n+1}\Delta_2^n\right). \tag{34}$$

Particularization of the above results to the case of absence of the vacuum gap is rather straightforward. However, unlike the standard implementation in Section 3.2.2, this time it inherently implies *imperfect* field restoration.

To sum up, the proposed DPS anti-cloak implementation allows only a *partial* recovery of the field in the inner spherical region $r < R_1$. In Section 4 below, we show some representative results, also in connection with sensor cloaking, and quantitatively assess the performance degradations with respect to standard implementations.

## 4. Representative numerical results

We now move on to illustrating some representative numerical results, obtained via the generalized Mie series in (10) and (11), so as to assess the analytical-based predictions in Section 3, within and beyond the ideal limit $\Delta_2, \Delta_3 \to 0$, and also in the more realistic case of lossy materials.

In particular, in all examples presented below, we consider a slight level of losses (loss-tangent=0.001) in the DPS media, and a higher level (loss-tangent=0.01) in the DNG media. As shown in our previous 2-D investigations [28,29], in the presence of losses, the "optimal" performance is not necessarily obtained in the ideal limit $\Delta_2, \Delta_3 \to 0$, and a parameter optimization is generally required. However, comprehensive parametric studies and/or exhaustive parameter optimizations are beyond the scope of the present prototype study, which is instead aimed at illustrating the basic effects and mechanisms. Accordingly, the results below were obtained by tweaking only a limited number of parameters, up to reaching a satisfactory response. This clearly leaves room for further improvements.

We start considering the standard cloak/anti-cloak implementation with the vacuum gap (cf. Section 3.2.1), i.e., the direct generalization of the cylindrical results in [28]. Figure 2 illustrates the field-map (real part of the electric field $x$-component) in the $y$-$z$ plane, for a plane-wave-excited configuration featuring $\varepsilon_2 = \varepsilon_1 = \varepsilon_0, \mu_2 = \mu_1 = \mu_0$, i.e., a DNG anti-cloak. Similar effects as for the cylindrical case are observed, with a very weak field intensity in the vacuum gap (which is thus *effectively cloaked*) and very weak exterior scattering, and yet the anti-cloak capability of restoring a cavity-type field in the inner (vacuum) spherical region $r < R_1$. As for the cylindrical case [28], qualitatively similar results (not shown for brevity) were observed for the alternative configuration featuring a DPS anti-cloak with a DNG inner sphere, as well as for an *epsilon-negative* anti-cloak with a *mu-negative* inner sphere (and vice versa).

For the same geometrical configuration, Fig. 3 shows the response obtained via an approximate DPS implementation of the anti-cloak (cf. Section 3.3). In this case, we fixed the constitutive parameters $\varepsilon_2 = -\varepsilon_0, \mu_2 = -\mu_0$, and limited ourselves to enforcing the condition (29) for $n = 1$ only, by exploiting the remaining parameter $R'_1$. We found that the arising transcendental equation has infinite solutions, and picked the smallest positive one $(R'_1 = 0.732 R_1)$. As it can be observed, in spite of the simplifications, the response attained is qualitatively comparable (in terms of cloaking and field restoration capabilities, as well as exterior scattering) with that exhibited by the standard (i.e., DNG anti-cloak) implementation.

Next, we considered the configuration without the vacuum gap (cf. Section 3.2.3), i.e., $R_3 = R_2$. As illustrated in Fig. 4, by enforcing the conditions in (25), it is now possible to quite faithfully restore in the inner spherical region $r < R_1$ the planar wavefronts of the impinging field, while still maintaining a very weak exterior scattering. Moreover, via the choice of the constitutive parameters in this region, it is also possible to control the field amplitude (cf. (26)). In particular, in our example, by choosing $\varepsilon_2 = 2\varepsilon_0$, we obtain a restored field twice as strong as the impinging one.

This naturally brings us to the illustration of the application to sensor cloaking. In this case, we considered a small, lossy inner spherical target of radius $R_1 = \lambda_0/8$ and permittivity $\varepsilon_1 = (4 + 0.4i)\varepsilon_0$, in order to mimic the sensor-loading effects. As in [29], we compactly parameterized the overall target "visibility" (to a far-field observer) via its total scattering cross section [38]

$$Q_s = \frac{\lambda_0^2}{2\pi} \sum_{n=1}^{\infty} (2n+1)\left(\left|a_{n,TE}^{(5)}\right|^2 + \left|a_{n,TM}^{(5)}\right|^2\right), \qquad (35)$$

and its "sensing" (absorption) capabilities via the time-averaged dissipated power

$$P_a = \frac{\omega \operatorname{Im}(\varepsilon_1)}{2} \int_0^{R_1} dr \int_0^{\pi} d\theta \int_0^{2\pi} d\phi \, r^2 \sin\theta \left|\boldsymbol{E}(r,\theta,\phi)\right|^2. \qquad (36)$$

For more direct understanding of the reductions/enhancements attainable, we normalized the above observables with respect to the corresponding reference values $Q_s^{(0)}$ and $P_a^{(0)}$ exhibited by the target free standing in vacuum, viz.,

$$\bar{Q}_s \equiv \frac{Q_s}{Q_s^{(0)}}, \quad \bar{P}_a \equiv \frac{P_a}{P_a^{(0)}}. \tag{37}$$

With reference to a standard DNG anti-cloak implementation, Fig. 5 illustrates response attainable, in terms of the above normalized observables, as a function of the parameters $\Delta_2$ and $\Delta_3$. Both observables turn out to exhibit broad dynamic ranges, which span from strong scattering/absorption reductions (with respect to the reference responses in vacuum) to values *higher* than those in vacuum; such *super-scattering/absorption* behaviors are typical of complementary media [39-41]. Qualitatively similar results, though with more compressed dynamic ranges, are observed in the case of DPS (approximate) anti-cloak implementation, illustrated in Fig. 6. As discussed in [29], a meaningful way of quantitatively assessing the sensor-cloaking performance is via a tradeoff curve which, for a given value of the scattering response yields the strongest absorption response achievable. Figure 7 compares such tradeoff curves (extracted from Figs. 5 and 6) for the DNG and DPS anti-cloak cases. As for the cylindrical case [29], we note that, varying $\Delta_2$ and $\Delta_3$, it is possible to span the entire range of cloak/anti-cloak interactions, going from a *cloak-prevailing* regime (with weak scattering/absorption) to an *anti-cloak-prevailing* one (with scattering and absorption comparable with or even higher than those in vacuum), and passing through an intermediate regime featuring relevant scattering reductions accompanied by sensible absorption levels.

As expectable, it can be observed that the DNG anti-cloak implementation consistently outperforms the approximate DPS one. For instance, assuming a targeted scattering reduction of -15 dB, the DNG anti-cloak implementation actually provides a *super-absorption* of 7.7 dB. The corresponding near-field response is shown in Fig. 8, from which it can be observed the very weak exterior scattering, and yet the power-coupling capabilities (as a reference, the response of the target alone free standing in vacuum is shown in Fig. 9). Nevertheless, for the same targeted scattering reduction, the approximate DPS configuration still provides an acceptable response (with an absorption only 1 dB below that of the free-standing target, and the corresponding near-field map shown in Fig. 10), whose deterioration with respect to the DNG case is fairly counterbalanced by a significantly simpler implementation.

In order to better understand the somehow counterintuitive behavior outlined above, it is insightful to look at the analytical structure of the *total absorption cross-section* [38]

$$Q_a = -\frac{\lambda_0^2}{2\pi} \sum_{n=1}^{\infty} (2n+1) \left[ \left|a_{n,TE}^{(5)}\right|^2 + \left|a_{n,TM}^{(5)}\right|^2 + \text{Re}\left(a_{n,TE}^{(5)} + a_{n,TM}^{(5)}\right) \right], \tag{38}$$

which is representative of the target absorption only in the ideal case of lossless cloak and anti-cloak. By comparison with the total scattering cross-section in (35), one notes that $Q_a$ *asymptotically* vanishes in the limit of zero scattering (i.e., $a_{n,TE,TM}^{(5)} \to 0$). Nevertheless, its vanishing trend may be *moderately slower* than that of $Q_s$, in view of the presence of the *linear* terms in (38). In other words, if zero scattering requires zero absorption, the ratio of absorption over scattering cross-sections may be unbounded [30]. This explains the possibility of achieving, outside the asymptotic regime, significant scattering reductions (with respect to the reference free-standing target) while maintaining sensible absorption levels.

As in [29], also shown in Fig. 6, as a reference, is the tradeoff curve pertaining to an imperfect cloak *alone* (i.e., without the anti-cloak) characterized by a standard (linear) transformation

$$r' = f(r) = \begin{cases} r, & r < R_1, \quad r > R_4, \\ R_4 \left( \dfrac{r - R_1 + \Delta_1}{R_4 - R_1 + \Delta_1} \right), & R_1 < r < R_4, \end{cases} \tag{39}$$

and same total thickness $(R_4 - R_1)$ as the above cloak/anti-cloak configurations. In this case, the curve is obtained by simply scanning the only parameter $(\Delta_1)$ available. As it can be observed, although the general trend looks similar to the above cases, the performance is considerably poorer, and allows a sensible scattering reduction (e.g., -15 dB) only at the expense of a severe curtail (-12 dB reduction in the absorption) of the sensing capabilities. As for the cylindrical case [29], this illustrates the important role played by the anti-cloak in the sensor cloaking mechanism.

## 5. Concluding remarks and hints for future research

In this paper, we have presented an analytical full-wave study of cloak/anti-cloak interactions in a 3-D *spherical* scenario, which extends some of our previous results obtained in the 2-D cylindrical case [28,29]. Our study, based on a generalized Mie-series approach, has confirmed for the more realistic 3-D scenario the intriguing field effects observed in the 2-D case, and has illustrated some new possibilities. In particular, we have shown the possibility of effectively cloaking an annular spherical region separating the cloak and the anti-cloak, while still being able to restore in an inner region a distorted version of the impinging field. Moreover, with the cloak and anti-cloak *directly contiguous*, we have shown the possibility of "perfectly" restoring the impinging field (controlling its amplitude), while maintaining a very weak exterior scattering, and its possible application to the sensor cloaking scenario. Finally, we have focused on the possibility of implementing *approximate* anti-cloaking effects via the use of DPS media only.

Our results constitute a further step towards the understating of cloak/anti-cloak interactions, and their fruitful exploitation in TO-based devices. In particular, the perspective of avoiding the use of DNG media looks very promising from the practical implementation viewpoint.

In this framework, current and future studies are aimed at exploring further parameter simplifications/reductions, based, e.g., on *spatially-invariant* [42] or *non-magnetic* [43,44] material parameters. Also of interest is a comparative study of the sensor-cloaking performance, against the alternative approaches [30-32]. In this framework, it is worth stressing, once again, that in this prototype study no particular effort was made to *exhaustively* optimize the parametric configurations pertaining to the application examples illustrated, and therefore we believe that significant room exists for further performance improvement.

**References**

[1] D. A. B. Miller, On perfect cloaking, Opt. Express 14 (2006) 12457–12466.

[2] F. Guevara Vasquez, G. W. Milton, D. Onofrei, Broadband exterior cloaking, Opt. Express 17 (2009) 14800–14805.

[3] F. Guevara Vasquez, G. W. Milton, D. Onofrei, Active exterior cloaking for the 2D Laplace and Helmholtz equations, Phys. Rev. Lett. 103 (2009) 073901.

[4] J. E. Ffowcs Williams, Review lecture: Anti-sound, Proc. R. Soc. A 395 (1984) 63–88.

[5] A. Alù, N. Engheta, Achieving transparency with plasmonic and metamaterial coatings, Phys. Rev. E 72 (2005) 016623.

[6] M. G. Silveirinha, A. Alù, N. Engheta, Cloaking mechanism with antiphase plasmonic satellites, Phys. Rev. B 78 (2008) 205109.

[7] B. Edwards, A. Alù, M. G. Silveirinha, N. Engheta, Experimental verification of plasmonic cloaking at microwave frequencies with metamaterials, Phys. Rev. Lett. 103 (2009) 153901.

[8] G. W. Milton, N. A. P. Nicorovici, On the cloaking effects associated with anomalous localized resonance, Proc. R. Soc. London A 462 (2006) 3027–3059.

[9] A. Greenleaf, M. Lassas, G. Uhlmann, Anisotropic conductivities that cannot be detected by EIT, Physiol. Meas. 24 (2003) 413–419.

[10] U. Leonhardt, Optical conformal mapping, Science 312 (2006) 1777–1780.

[11] J. B. Pendry, D. Schurig, D. R. Smith, Controlling electromagnetic fields, Science 312 (2006) 1780–1782.

[12] D. Schurig, J. J. Mock, B. J. Justice, S. A. Cummer, J. B. Pendry, A. F. Starr, D. R. Smith, Metamaterial electromagnetic cloak at microwave frequencies, Science 314 (2006) 977–980.

[13] W. Cai, U. K. Chettiar, A. V. Kildishev, V. M. Shalaev, Optical cloaking with metamaterials, Nature Photonics 1 (2007) 224.

[14] A. V. Kildishev, W. Cai, U. K. Chettiar, V. M. Shalaev, Transformation optics: approaching broadband electromagnetic cloaking, New J. Phys. 10 (2008) 115029.

[15] M. Yan, W. Yan, M. Qiu, Invisibility cloaking by coordinate transformation, in: E. Wolf (Ed.), Progress in Optics, vol. 52, Elsevier, Hungary, 2009, pp. 261–304.

[16] A. Håkansson, Cloaking of objects from electromagnetic fields by inverse design of scattering optical elements, Opt Express. 15 (2007) 4328–34.

[17] P. Alitalo, O. Luukkonen, L. Jylha, J. Venermo, S. A. Tretyakov, Transmission-line networks cloaking objects from electromagnetic fields, IEEE Trans. Antennas Propagat. 56 (2008) 416–424.

[18] P. Alitalo, F. Bongard, J.-F. Zurcher, J. Mosig, S. Tretyakov, Experimental verification of broadband cloaking using a volumetric cloak composed of periodically stacked cylindrical transmission-line networks, Appl. Phys. Lett. 94 (2009) 014103.

[19] A. Alù, Mantle cloak: Invisibility induced by a surface, Phys. Rev. B 80 (2009) 245115.

[20] A. Alù, N. Engheta, Plasmonic and metamaterial cloaking: physical mechanisms and potentials, J. Opt. A: Pure Appl. Opt. 10 (2008) 093002.

[21] P. Alitalo, H. Kettunen, S. Tretyakov, Cloaking a metal object from an electromagnetic pulse: A comparison between various cloaking techniques, J. Appl. Phys. 107 (2010) 034905.

[22] Z. Ruan, M. Yan, C. W. Neff, M. Qiu, Ideal cylindrical cloak: Perfect but sensitive to tiny perturbations, Phys. Rev. Lett. 99, 113903 (2007).

[23] B. Zhang, H. S. Chen, B. I. Wu, Y. Luo, L. X. Ran, J. A. Kong, Response of a cylindrical invisibility cloak to electromagnetic waves, Phys. Rev. B 76 (2007) 121101.

[24] H. S. Chen, B. I. Wu, B. Zhang, J. A. Kong, Electromagnetic wave interactions with a metamaterial cloak, Phys. Rev. Lett. 99 (2007) 063903.

[25] Zhang, H. Chen, B. I. Wu, J. A. Kong, Extraordinary surface voltage effect in the invisibility cloak with an active device inside, Phys. Rev. Lett. 100 (2008) 063904.

[26] H. Chen, X. Luo, H. Ma, C. T. Chan, The anti-cloak, Opt. Express 16 (2008) 14603–14608.

[27] R. W. Ziolkowski, N. Engheta, Introduction, history and fundamental theories of double-negative (DNG) metamaterials, in: N. Engheta, R. W. Ziolkowski (Eds.), Metamaterials: Physics and Engineering Explorations, Wiley-IEEE Press, Piscataway, NJ, 2006, pp. 5-41.

[28] G. Castaldi, I. Gallina, V. Galdi, A. Alù, N. Engheta, Cloak/anti-cloak interactions, Opt. Express 17 (2009) 3101–3114.

[29] G. Castaldi, I. Gallina, V. Galdi, A. Alù, N. Engheta, Power scattering and absorption mediated by cloak/anti-cloak interactions: A transformation-optics route towards invisible sensors, J. Opt. Soc. Am. B 27 (2010) 2132–2140.

[30] A. Alù, N. Engheta, Cloaking a sensor, Phys. Rev. Lett. 102 (2009) 233901.

[31] Z. Ruan, S. Fan, Temporal Coupled-mode theory for Fano resonance in light scattering by a single obstacle, J. Phys. Chem. C 114 (2010) 7324–7329.

[32] A. Greenleaf, Y. Kurylev, M. Lassas, G. Uhlmann, Cloaking a sensor via transformation optics, Phys. Rev. E 83 (2011) 016603.

[33] Y. Luo, H. Chen, J. Zhang, L. Ran, J. A. Kong, Design and analytical full-wave validation of the invisibility cloaks, concentrators, and field rotators created with a general class of transformations, Phys. Rev. B 77 (2008) 125127.

[34] I. Gallina, G. Castaldi, V. Galdi, A. Alù, N. Engheta, General class of metamaterial transformation slabs Phys. Rev. B 81 (2010) 125124.

[35] R. F. Harrington, Time-Harmonic Electromagnetic Fields, IEEE Press-Wiley Interscience, Piscataway, NJ, 2001.

[36] M. Abramowitz, I. A. Stegun, Handbook of Mathematical Functions, Ninth printing, Dover, New York, 1970.

[37] M. Rahm, D. Schurig, D. A. Roberts, S. A. Cummer, D. R. Smith, J. B. Pendry, Design of electromagnetic cloaks and concentrators using form-invariant coordinate transformations of Maxwell's equations, Photon. Nanostruct. 6 (2008) 87–95.

[38] J. A. Stratton, Electromagnetic Theory, McGraw-Hill, New York, 1941.

[39] N. A. Nicorovici, R. C. McPhedran, G. W. Milton, Optical and dielectric properties of partially resonant composites, Phys. Rev. B 49 (1994), 8479–8482.

[40] T. Yang, H. Chen, X. Luo, H. Ma, Superscatterer: Enhancement of scattering with complementary media, Opt. Express 16 (2008) 18545.

[41] J. Ng, H. Chen, C. T. Chan, Metamaterial frequency-selective superabsorber, Opt. Lett. 34, (2009) 644–464.

[42] Y. Luo, J. Zhang, H. Chen, S. Xi, B.-I. Wu, Cylindrical cloak with axial permittivity/permeability spatially invariant, Appl. Phys. Lett. 93 (2008) 033504.

[43] W. Cai, U. K. Chettiar, A. V. Kildishev, V. M. Shalaev, G. W. Milton, Nonmagnetic cloak with minimized scattering, Appl. Phys. Lett. 91 (2007) 111105.

[44] G. Castaldi, I. Gallina, V.Galdi, Nearly perfect nonmagnetic invisibility cloaking: Analytic solutions and parametric studies, Phys. Rev. B 80 (2009) 125116.

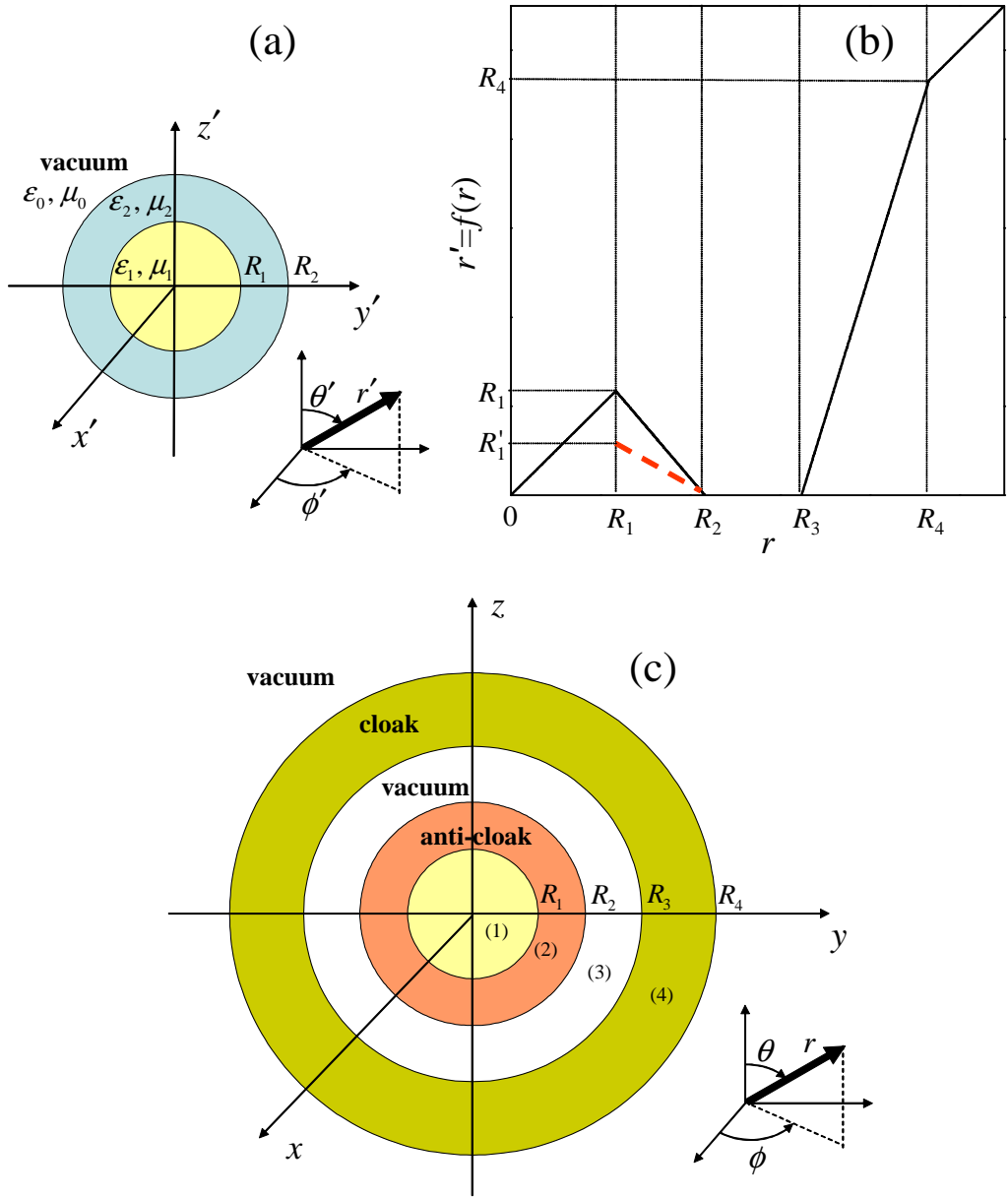

**Figure 1** – (Color online) Problem geometry. (a) Two-layer piecewise-homogeneous, isotropic spherical configuration in the fictitious auxiliary space $(x', y', z')$ (and associated $(r', \theta', \phi')$ spherical coordinate system). (b) Piecewise-linear radial coordinate transformation in (2) (in the limit $\Delta_2, \Delta_3 \to 0$), with the red-dashed part illustrating the discontinuous case considered in Section 3.3. (c) General cloak/anti-cloak (with vacuum gap) configuration in the actual physical space $(x, y, z)$ (and associated $(r, \theta, \phi)$ spherical coordinate system).

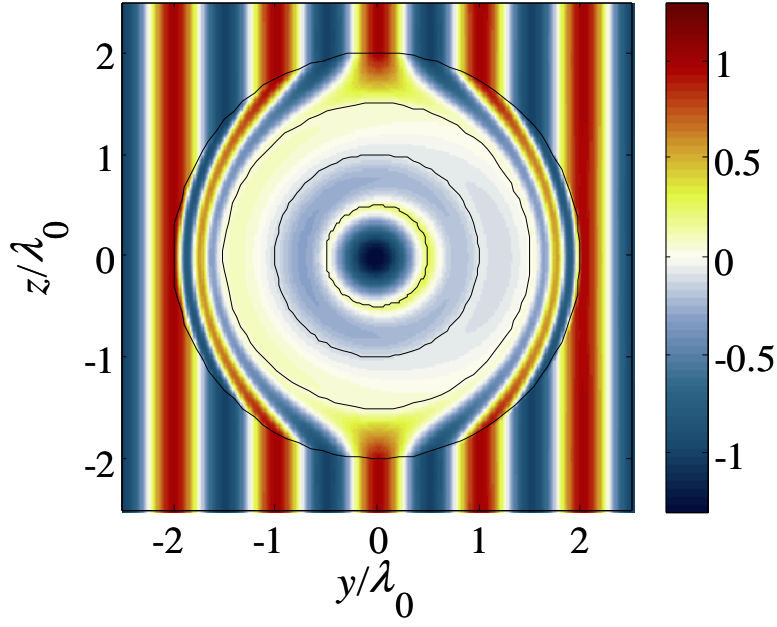

**Figure 2** – (Color online) Representative field-map (real part of electric field *x*-component in the *y*-*z* plane) for the plane-wave-excited configuration in Fig. 1(c) with $\varepsilon_2 = \varepsilon_1 = \varepsilon_0$, $\mu_2 = \mu_1 = \mu_0$ (i.e., DNG anti-cloak), $R'_1 = R_1 = \lambda_0/2$, $R_2 = \lambda_0$, $R_3 = 3\lambda_0/2$, $R_4 = 2\lambda_0$, $\Delta_2 = R_2/20$, and $\Delta_3 = R_3/100$. Here, and henceforth, loss-tangent values of 0.001 and 0.01 are consistently assumed for DPS and DNG media, respectively.

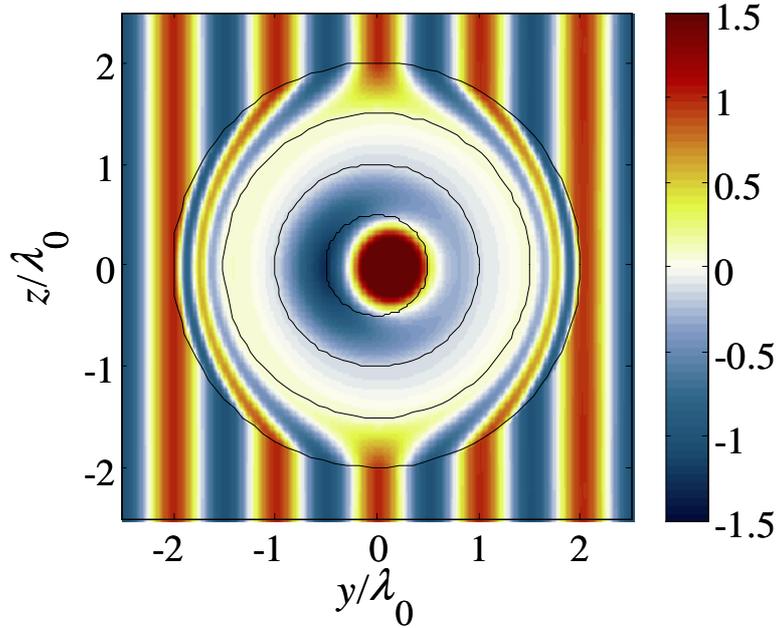

**Figure 3** – (Color online) As in Fig. 2, but for approximate DPS implementation of the anti-cloak (cf. Section 3.3), with $\varepsilon_2 = -\varepsilon_0$, $\mu_2 = -\mu_0$, and $R'_1 = 0.732 R_1$.

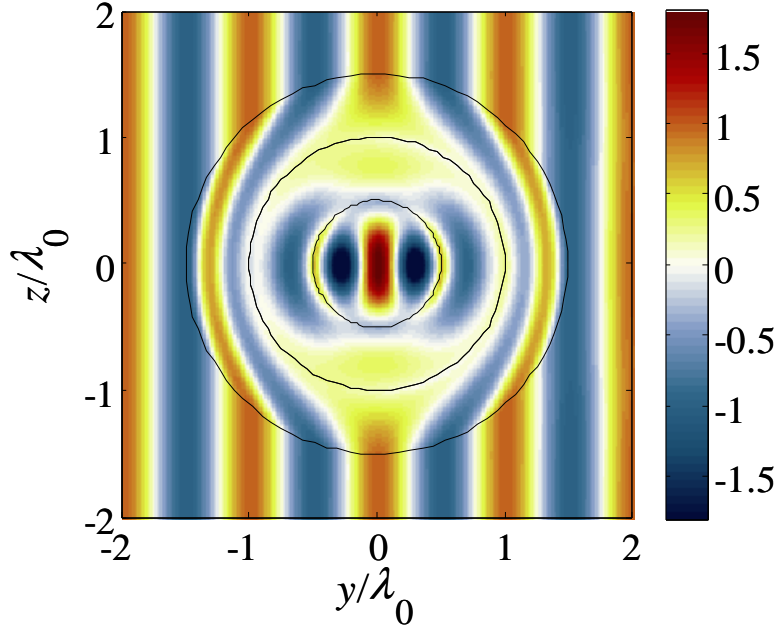

**Figure 4** – (Color online) As in Fig. 2 (i.e., DNG anti-cloak), but in the absence of the vacuum gap, and with $\varepsilon_2 = \varepsilon_1 = 2\varepsilon_0$, $\mu_2 = \mu_1 = 2\mu_0$, $R_3 = R_2 = \lambda_0$, $R_4 = 3\lambda_0/2$, $\Delta_2 = R_2/10$, $\Delta_3 = R_3/20$.

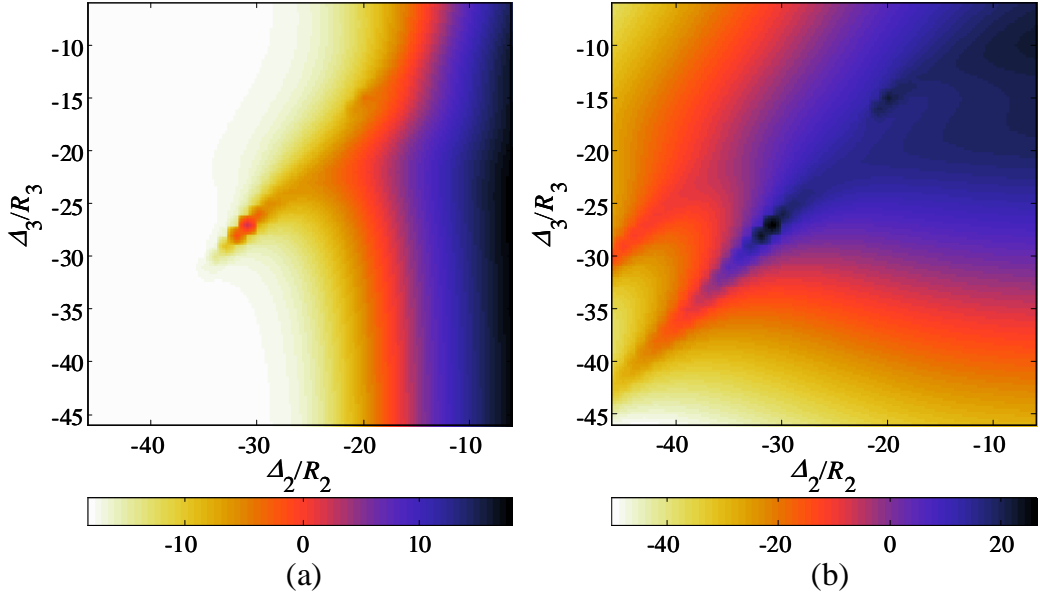

**Figure 5** – (Color online) Geometry as in Fig.1(c), but in the absence of the vacuum gap, with $\varepsilon_1 = (4+0.4i)\varepsilon_0$, $\varepsilon_2 = 4\varepsilon_0$, $\mu_2 = \mu_1 = \mu_0$ (i.e., DNG anti-cloak), $R_1' = R_1 = \lambda_0/8$, $R_2 = R_3 = \lambda_0/2$, and $R_4 = \lambda_0$. (a) Normalized total scattering cross-section (cf. (37)) in dB scale, as a function of $\Delta_2/R_2$ and $\Delta_3/R_3$. (b) Corresponding normalized time-averaged dissipated power (cf. (37)).

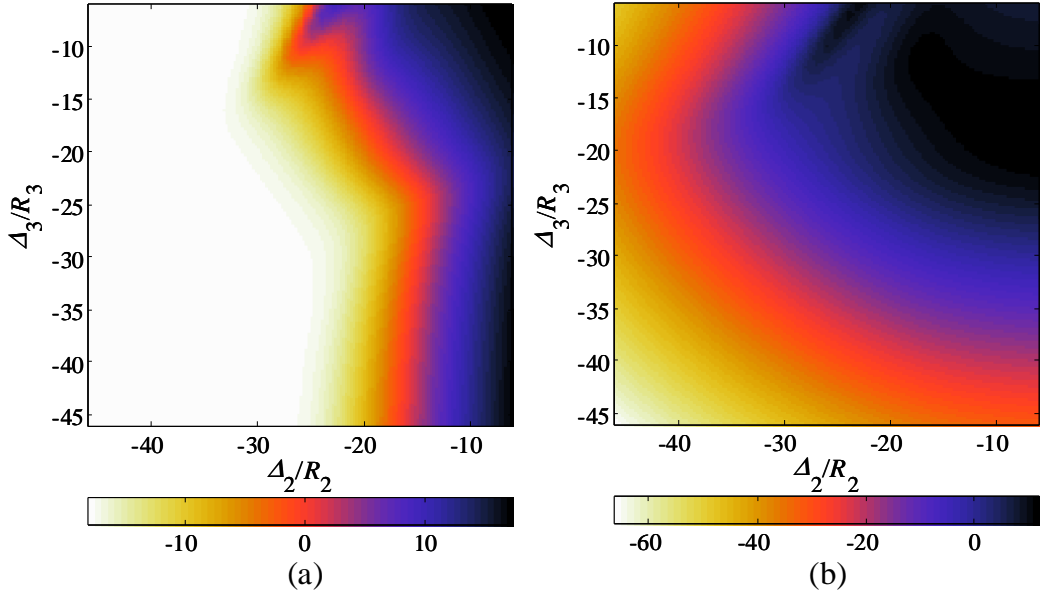

**Figure 6** – (Color online) As in Fig. 5, but for an approximate DPS implementation of the anti-cloak, with $\varepsilon_2 = -4\varepsilon_0$, $\mu_2 = -\mu_0$, and $R'_1 = 0.56 R_1$.

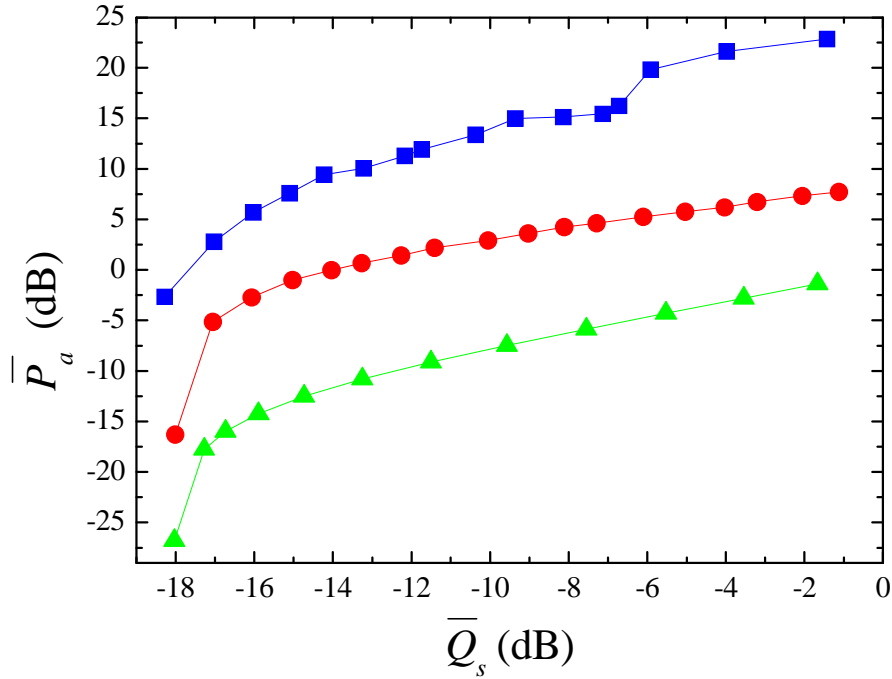

**Figure 7** – (Color online) Sensor-cloaking tradeoff curves pertaining to the DNG (squares) and approximate DPS anti-cloak (circles) configurations (cf. Figs. 5 and 6, respectively). Also shown (triangles), as a reference, is the corresponding curve pertaining to an imperfect cloak configuration (cf. (39)).

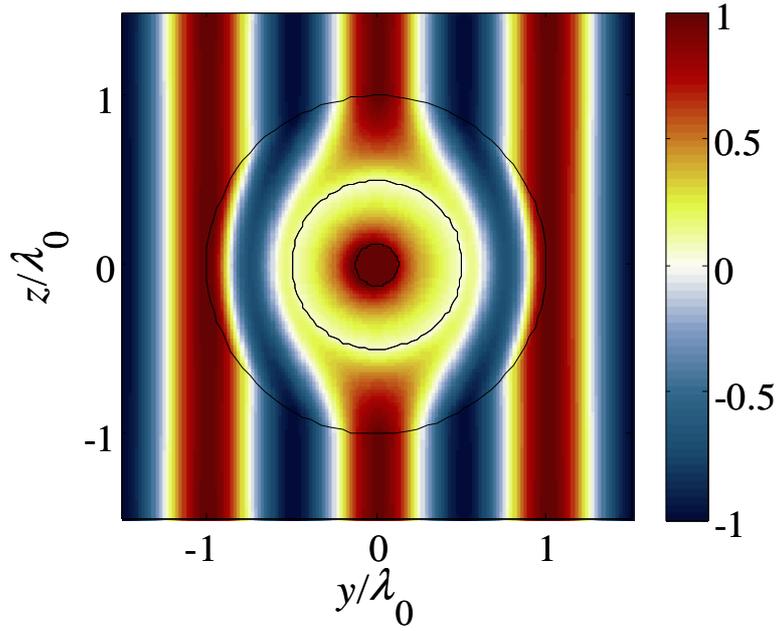

**Figure 8** – (Color online) Representative field map pertaining to the cloaked sensor with DNG anti-cloak (cf. Fig. 5), with $\Delta_2 = 0.125 R_2$ and $\Delta_3 = 0.045 R_3$, featuring $\bar{Q}_s = -15\text{dB}$ and $\bar{P}_a = 7.7\text{dB}$.

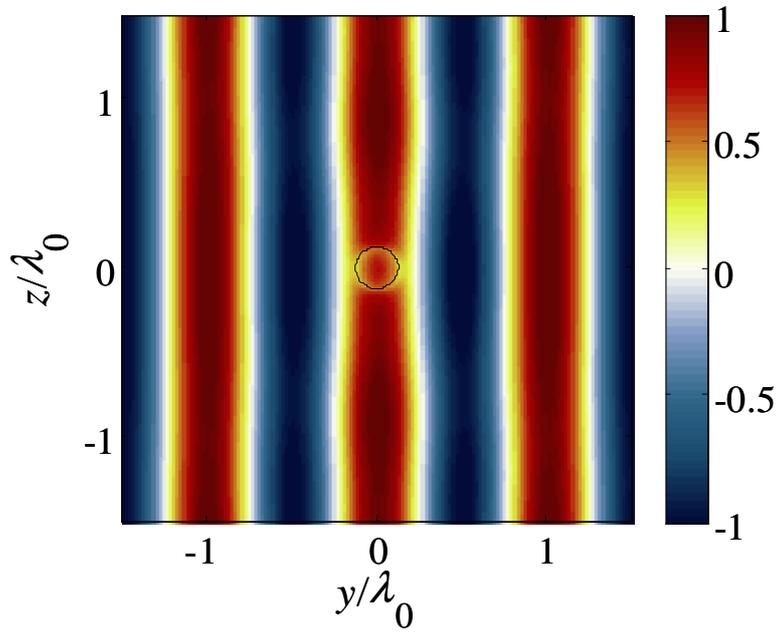

**Figure 9** – (Color online) As in Fig. 8, but for the target free standing in vacuum.

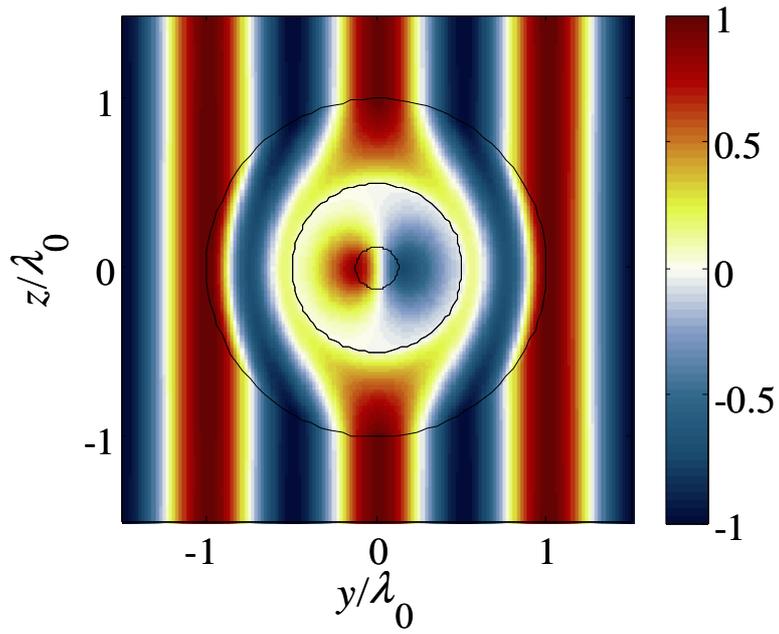

**Figure 10** – (Color online) As in Fig. 8, but for approximate DPS implementation of the anti-cloak, with $\Delta_2 = 0.158 R_2$ and $\Delta_3 = 0.0315 R_3$, featuring $\bar{Q}_s = -15\text{dB}$ and $\bar{P}_a = -1\text{dB}$.